# Choix stratégiques de la firme et contrôle financier

**Jean-Claude JUHEL**

L'observation du fonctionnement des organisations productives montre que les caractéristiques d'un métier, adossées par nature à un environnement technologique donné, déterminent la **combinaison productive** mise en oeuvre par le décideur, et, la **structure du cycle d'exploitation** qui lui est afférent. L'exploitation d'une grande surface de distribution ou d'un chantier naval impose des spécifications technologiques et organisationnelles adaptées et propres au niveau de compétitivité de la branche.

Le choix de la combinaison productive et le choix de la structure du cycle d'exploitation obèrent les conditions dans lesquelles la trésorerie de la firme va évoluer. De nouveaux outils de contrôle financier - l'effet de levier de trésorerie et le surplus de trésorerie d'exploitation - donnent à l'entrepreneur les informations utiles à l'efficience des choix stratégiques de la firme.

## I - Le choix de la combinaison productive.

La nature de la combinaison productive détermine la **structure des charges et donc le risque de défaillance** assumé par l'entreprise.

L'adaptation du modèle, bien connu, « profit-coûts-volume » à la gestion de la sécurité de l'entreprise fournit un outil d'analyse du **degré de liquidité** d'une combinaison productive : **le seuil de solvabilité**. Mais on peut aller au delà d'un simple approfondissement du seuil de rentabilité **par une généralisation du concept.** L'application le la notion d'élasticité à l'analyse de la sensibilité de la trésorerie permet de définir l'**effet de levier de trésorerie**.

### A - Le seuil de solvabilité.

Si on se limite à l'étude des seuls flux d'exploitation, on constate que le degré de liquidité d'une entreprise dépend de la façon dont les flux de trésorerie qui leur sont associés se présentent les uns par rapport aux autres dans le temps. Le seuil de solvabilité, pour une période déterminée ou pour un cycle d'exploitation donné, est le volume d'activité pour lequel la trésorerie afférente est nulle. En deçà de ce seuil la firme est dans une zone d'insolvabilité, au-delà, disparaît le risque d'une crise de trésorerie. Expression des décalages entre les flux d'exploitation et les flux de trésorerie, le seuil de solvabilité est un indicateur de sécurité permettant de mesurer le risque d'insolvabilité encouru par l'entreprise[1].

Cet indicateur intègre deux données fondamentales.

D'une part, tous les coûts fixes ne sont pas « décaissables ». Ce sont les « charges calculées », sans influence sur la trésorerie immédiate, bien qu'à terme elles jouent sur le degré de solvabilité.

D'autre part, la structure temporelle des coûts décaissables conditionnent la solvabilité de la firme. Dans la mesure où la contrainte des coûts décaissables sur la trésorerie ne

[1] Juhel J.-C., « Le seuil de solvabilité, instrument d'analyse financière et modèle de prise de décisions » Revue Française de Comptabilité, n° 262, décembre 1994, pp. 60-69.

s'exerce qu'à l'instant de leur déboursement, il paraît opportun de distinguer la *« liquidité »* d'une opération de sa *« solvabilité »* :

1° Nous dirons qu'une opération est *« liquide »*, au moment où tous les coûts décaissables, quelle que soit leur échéance au cours de la période, sont couverts par le revenu de l'activité ; on peut nommer ce point-mort de trésorerie, *« seuil de liquidité »*, pour signifier que l'opération garantit à un certain niveau de production une trésorerie positive, tout en ignorant la date où elle le deviendra.

2° Nous dirons qu'une opération est *« solvable »*, au moment où la trésorerie de l'opération devient effectivement positive ; on peut nommer ce second point-mort de trésorerie, *« point-mort d'encaisse »* ou *« seuil de solvabilité »*, pour signifier que l'entreprise honore tous ses engagements à compter de cette date.

Le seuil de liquidité, par analogie au seuil de rentabilité s'écrit :

$$\textbf{Seuil de liquidité} = \frac{\textbf{Coûts fixes décaissables}}{\textbf{Marge unitaire sur coûts variables}}$$

Le seuil de liquidité devient seuil de solvabilité par intégration des décalages éventuels des flux de trésorerie par rapport aux flux d'exploitation[2]. Cette intégration s'effectue en quatre étapes :

*1°- La modulation des charges de structure décaissables* ; c'est à dire leur étalement naturel dans le temps plus ou moins maîtrisé par le décideur, et, le cas échéant, leur paiement différé par crédit-fournisseur.

2°- *L'anticipation des charges variables* ; autrement dit l'incidence du volume des stocks de consommations intermédiaires sur les flux d'encaisse.

3°- *Le décalage des sorties de fonds* liées aux coûts variables, dû au crédit-fournisseur.

4°- *Le décalage de l'encaissement du chiffre d'affaires* dû au crédit-client.

En outre, si l'activité est saisonnière, on peut intégrer dans le modèle l'influence de la saisonnalité des flux d'exploitation sur les flux de trésorerie.

Calculé rétrospectivement, le seuil de solvabilité permet de préciser la marge de sécurité présentée par les relations « encaisse-coûts-volume» et ainsi de mesurer le risque engendrer par les activités de production ; calculé prospectivement, il permet de préparer les décisions ultérieures de gestion dans une perspective volontariste, en vue de diminuer les risques de défaillance. Le modèle reproduit les décalages entre flux d'exploitation et flux de trésorerie et détermine leur impact sur la solvabilité de l'entreprise. En l'absence de décalage, le seuil de solvabilité est égal au seuil de liquidité.

### a) La modulation des charges de structure décaissables.

Le paiement des charges de structure s'échelonne, par la nature des choses, tout au long de l'exercice selon un rythme propre à chaque entreprise. Le responsable peut déplacer plus ou moins facilement ces décaissements, et obtenir, en outre, grâce au crédit des fournisseurs, un report des échéances. Cette *« modulation »* des coûts fixes décaissables réduit le montant à décaisser en début de période et améliore la situation de trésorerie. Le seuil de solvabilité s'écrit alors :

$$\textbf{Seuil de solvabilité} = \frac{\textbf{Coûts fixes décaissables - Coûts fixes « modulés » à décaisser}}{\textbf{Marge unitaire sur coûts variables}}$$

[2] Les flux de T.V.A. sont des flux de trésorerie qui jouent sur la liquidité et la solvabilité selon un mécanisme très particulier dont nous avons présenté une analyse par ailleurs - voir : Juhel J.-C., « L'effet T.V.A. : variable financière de gestion », Revue Française de Comptabilité, n° 217, novembre 1990, pp.74-82. Nous n'en tiendrons pas compte dans la présente étude.

Cependant, la modulation n'exerce totalement son effet sur la solvabilité que si le décaissement des « charges de structure modulables » est repoussé après la vente de la production correspondant au seuil de solvabilité initial, c'est à dire avant modulation.

### b) L'anticipation des charges variables : le poids des stocks de consommations intermédiaires sur la trésorerie.

Généralement les consommations intermédiaires stockables nécessaires à la production doivent être acquises avant le début du cycle d'exploitation. L'entrepreneur *« anticipe »* donc certains coûts variables et constitue ainsi des stocks de matières premières ou de marchandises dès avant le début du cycle et les supportera tout au long du programme de production. Les charges variables ainsi anticipées pèsent sur la solvabilité de l'entreprise. Leur montant s'ajoute aux coûts fixes décaissables en début de période. Le seuil de solvabilité devient après anticipation des charges variables concernées :

$$\textbf{Seuil de solvabilité} = \frac{\textbf{Coûts fixes décaissables + Coûts variables anticipés}}{\textbf{Marge unitaire}}$$

La constitution de stocks de consommations intermédiaires, par l'anticipation de frais variables qu'elle représente, grève lourdement la trésorerie. Il convient de préciser que l'anticipation des charges variables se présente selon **deux cas de figure** qui peuvent d'ailleurs coexister :

- **1er cas** - l'anticipation porte sur l'ensemble des charges variables des premières unités produites pendant une certaine durée du cycle d'exploitation ; cette contrainte provoque un simple décalage de la courbe des charges d'activité ; les coûts variables anticipés s'ajoutent aux coûts fixes ; la marge unitaire ne change pas. La sous-traitance et la distribution correspondent, par exemple, à cette situation.
- **2ème cas** - l'anticipation ne porte que sur tout ou partie d'un type de coûts variables ; par exemple, il est nécessaire d'acquérir toute la matière première d'une opération avant le début du cycle d'exploitation (le produit est saisonnier, ou cette condition est imposée par l'approvisionnement). Les coûts variables anticipés gonflent les coûts fixes ; la marge unitaire diminue d'autant ; la courbe des charges d'activité n'est pas décalée. La marge unitaire sur coûts variables doit donc être recalculée.

### c) Les flux de coûts variables décalés par le crédit-fournisseur.

Le *crédit-fournisseur* sur charges variables permet d'améliorer la situation de trésorerie. En particulier, le report des sorties de fonds correspondantes modère ou annule leur anticipation. Dans tous les cas le montant du crédit-fournisseur, ou *« coûts variables décalés »* se déduit des coûts fixes décaissables en début de période. La valeur des coûts variables décalés par le crédit-fournisseur se calcule simplement :

**= Coûts variables unitaires (FS) × Durée du crédit (mois) × Ventes mensuelles (unités).**

Il existe cependant une **limite** à la valeur du seuil. **Le crédit-fournisseur ne joue sur la situation de trésorerie que si le seuil de solvabilité est supérieur à la production couvrant les frais anticipés décaissables.** Au-dessous, en effet, les charges décaissables avant le début de période deviennent le facteur déterminant, et **la valeur du seuil** devient donc égale au rapport :

$$\textbf{Seuil de solvabilité} = \frac{\textbf{Coûts anticipés décaissables + Encaissements décalés}^{3}}{\textbf{Prix de vente unitaire}}$$

### d) L'encaissement décalé par le crédit-client.

Le ***crédit-client***, en repoussant dans le temps les encaissements, augmente considérablement le risque d'insolvabilité. Son montant s'ajoute à celui des coûts fixes décaissables en début de période. La valeur de l'encaissement décalé se calcule de la manière suivante :

$$= \textbf{Prix de vente unitaire (FS)} \times \textbf{Durée du crédit (mois)} \times \textbf{Ventes mensuelles (unités)}$$

Le seuil de solvabilité tend vers une **limite** lorsque le total des charges décaissables plafonne avant la fin du cycle, ce qui peut se produire sous l'effet du décalage entre flux d'encaissements et flux de décaissements. Ce seuil particulier est :

$$\textbf{Seuil de solvabilité} = \frac{\textbf{Charges totales décaissables + Encaissements décalés}}{\textbf{Prix de vente unitaire}}$$

En résumé, **le seuil de solvabilité d'une opération,** avant la prise en compte de l'influence de la saisonnalité sur les flux d'exploitation est égal au **rapport** entre :

d'une part,

- les coûts fixes décaissables, (+), **CFD**,
- la modulation des coûts fixes décaissables, (-), **MCFD**
- les coûts variables anticipés, (+), **CVA**,
- les coûts variables décalés, (-), **CVD**,
- l'encaissement décalé, (+), **ED**,

et, d'autre part,

- la marge unitaire sur coûts variables, **MUSCV**,

soit :

$$\textbf{Seuil de solvabilité} = \frac{\textbf{CFD - MCFD + CVA - CVD + ED}}{\textbf{MUSCV}}$$

**Remarque : l'influence de la saisonnalité des flux d'exploitation sur le seuil de solvabilité.** Dans le cas où les flux d'exploitation subissent une influence non négligeable de la saisonnalité de l'activité, il convient de déterminer le seuil de solvabilité à l'aide de la représentation graphique des séries chronologiques observées ou calculées de ces flux. En effet, la relation présentée suppose évidemment une évolution régulière des variables de l'exploitation.

---

[3] Cette relation fait intervenir, le cas échéant, le flux des encaissements décalés par le crédit-client.

Le seuil de solvabilité donne donc simplement et rapidement le volume de production pour lequel, la date à laquelle, et les conditions dans lesquelles la trésorerie deviendra effectivement positive. Mais pour parvenir à une meilleure mesure de la sensibilité de la trésorerie on peut proposer une généralisation de cet outil. Le risque d'insolvabilité dépend du niveau et de la nature des coûts fixes, de la marge unitaire sur coûts variables et du volume de production vendue. L'effet de levier de trésorerie rend compte de l'interaction de ces paramètres.

### B - L'effet de levier de trésorerie.

Nous allons montrer que pour mesurer le risque d'insolvabilité, on peut évaluer la relation qui existe entre la variation relative de la capacité d'autofinancement ou « trésorerie virtuelle »[4], et, d'une part, la variation relative de la production vendue, et, d'autre part, la variation relative de la marge unitaire sur coûts variables. Cette application du *concept d'élasticité* à l'analyse de la sensibilité de la trésorerie met en évidence les contraintes imposées par la combinaison productive retenue - capacité de production et structure des coûts - sur la liquidité de l'entreprise. Ainsi, deux phénomènes interdépendants conditionnent le risque d'insolvabilité :

- l'élasticité de la trésorerie par rapport à la variation du volume d'activité, et,
- l'élasticité de la trésorerie par rapport à la variation de la marge unitaire sur coûts variables.

#### a) L'élasticité de la trésorerie par rapport au volume d'activité : $E_{T/Q}$.

L'élasticité de la trésorerie par rapport par rapport au volume d'activité est le rapport entre la variation relative de la trésorerie et la variation relative de la production vendue. Ainsi, une élasticité de 2 signifie qu'une augmentation de 1 % de la production vendue entraîne une augmentation de 2 % de la trésorerie virtuelle.

Si l'on écrit que :

« T » est la trésorerie,
« Q » est le volume vendu,

l'élasticité de la trésorerie par rapport au volume d'activité s'exprime :

$$E_{T/Q} = \frac{\Delta T / T}{\Delta Q / Q}$$

Si l'on note que la trésorerie, T, est égale à :

$$T = [(p - v)\, Q] - F$$

où

« p » est le prix de vente unitaire,

[4] Du point de vue de la liquidité de la firme, la capacité d'autofinancement représente la variation potentielle de la trésorerie en fin de période. Avec une trésorerie nulle en début de période, la capacité d'autofinancement est la « *trésorerie virtuelle* » de fin de période. L'écart entre l'encaisse virtuelle et l'encaisse réelle tient aux intervalles temporels entre les flux d'exploitation et les flux de trésorerie.

« v » est le coût variable unitaire,
« F » est le montant des charges de structure,
et, « Q », les quantités au niveau desquelles l'élasticité est calculée,

et, si $m = p - v$ désigne la marge unitaire sur coûts variables, l'élasticité s'écrit alors :

$$E_{T/Q} = \frac{(p - v)\,Q}{[(p - v)\,Q] - F} = \frac{mQ}{mQ - F}$$

$$\mathbf{E_{T/Q} = \frac{Q}{Q - F/m}}$$

Cette relation permet de calculer l'effet de levier de la trésorerie pour une valeur quelconque de la production vendue, pour une marge donnée et pour un montant de charges fixes donné. L'élasticité varie naturellement à chaque niveau de production.

Mais tous les flux d'exploitation ne se transforment pas en flux de trésorerie. Les charges calculées essentiellement, à savoir les amortissements et les provisions de la période[5], sont sans influence sur la trésorerie immédiate bien qu'à terme par le biais des exigences de la protection et du renouvellement du patrimoine elles affectent le degré de liquidité. Deux indicateurs sont donc à prendre en considération : *l'élasticité de la trésorerie à terme* ou « effet de levier d'exploitation » - appelé également « effet de levier opérationnel » - et, *l'élasticité de la trésorerie immédiate* que l'on pourrait dénommer par analogie « effet de levier d'encaisse ». Dans le premier cas les coûts fixes « **F** » englobent les charges calculées ; dans le deuxième cas, ils ne comprennent que les charges décaissables.

Effet de levier d'exploitation et effet de levier d'encaisse mesurent la sensibilité de la trésorerie à toute variation du volume des ventes, pour un niveau de production déterminé et pour une marge unitaire donnée. Plus le coefficient de levier est fort, plus le degré de sensibilité de la trésorerie est élevé et attache à toute évolution de l'activité décidée ou subie, un risque important de détérioration ou d'amélioration soit de la liquidité à terme (ou rentabilité), soit de la liquidité immédiate (ou solvabilité potentielle), soit encore des deux. Mais quelle que soit la nature de l'entreprise ou de la combinaison productive l'effet d'amplification a la même configuration bien que lié au montant des coûts fixes et au volume d'activité. En effet :

$$\text{pour} \quad mQ = F \quad \Rightarrow \quad E_{T/Q} = \frac{mQ}{mQ - F} \rightarrow \infty$$

$$\text{donc} \quad \Rightarrow \quad (mQ) - F = 0$$

$$\text{d'où} \quad \mathbf{Q^* = F / m}$$

ou valeur de la production pour laquelle la marge sur coûts variables couvre les coûts fixes. Si **F** représente l'ensemble des charges de structure, on remarquera que **F/m**, la production critique, est l'expression du seuil de rentabilité. La rentabilité étant le gage de la liquidité à terme, le seuil de rentabilité peut être nommé « *seuil de liquidité à terme* » ou valeur de la production à partir de laquelle la trésorerie nette est susceptible de devenir positive après s'être assuré du maintien de la valeur du capital[6]. Par contre lorsque **F** ne désigne que les charges de structure décaissables la valeur de ce rapport est un « *seuil de liquidité immédiate* » ou valeur de la production à partir de laquelle la trésorerie nette est susceptible de devenir positive[7]. Le graphique suivant montre le comportement général du coefficient d'élasticité par rapport à la production.

---

[5] Il en est de même des produits « fixes » non encaissables, tels que « les transferts de charges » lorsqu'il s'agit de transferts à un compte de bilan.

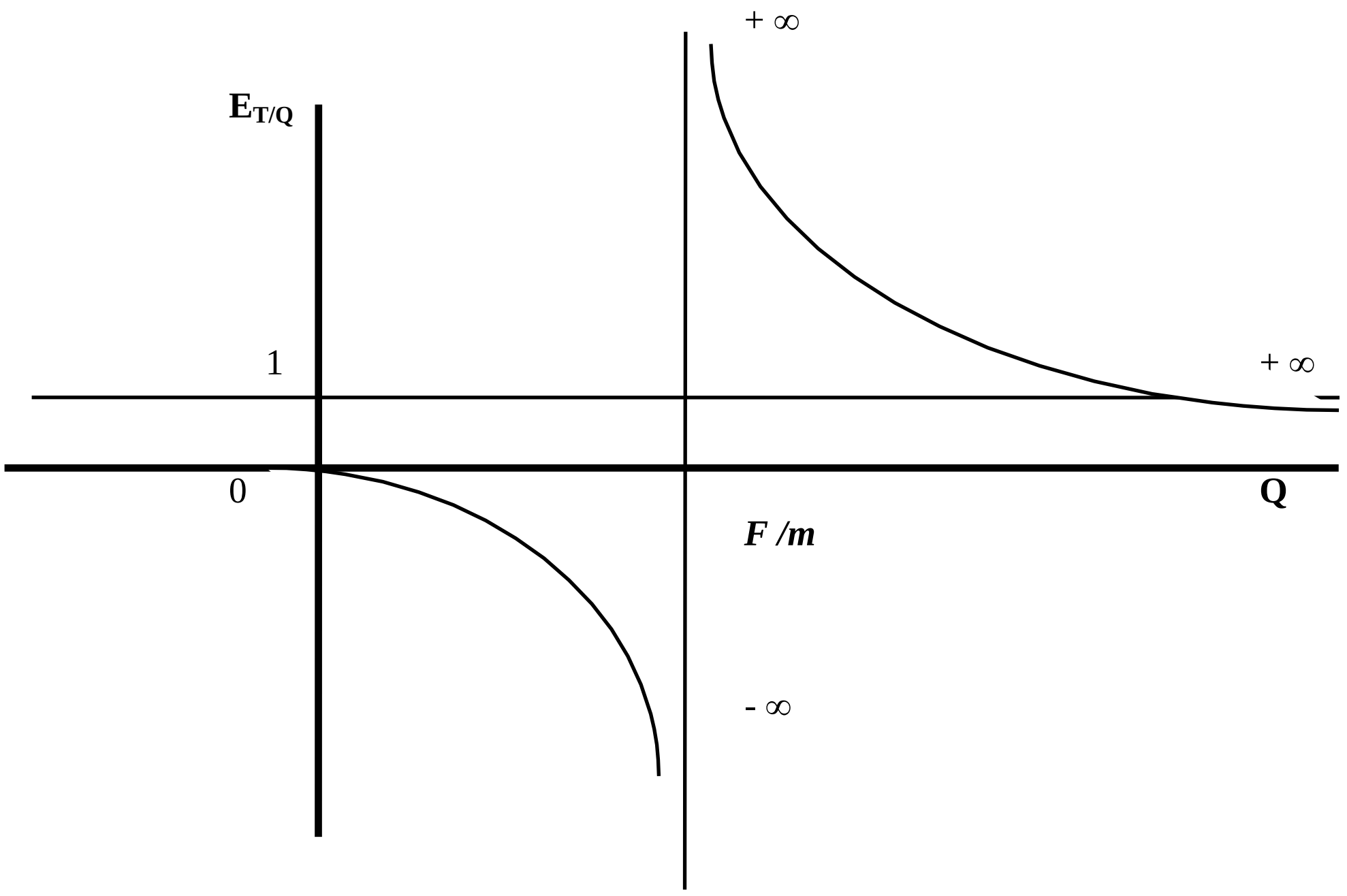

Les valeurs que peut prendre l'élasticité en fonction de la production sont résumées dans le tableau suivant :

| | | | | | | | | |
|---|---|---|---|---|---|---|---|---|
| Q | 0 | ½ F/m | 2/3 F/m | F/m | F/m | 2 F/m | 3 F/m | ∞ |
| E | 0 | -1 | -2 | - ∞ | + ∞ | 2 | 3/2 | 1 |

Aux seuils de liquidité immédiate et à terme, **Q***, lorsque la marge totale couvre les coûts fixes afférents, l'élasticité tend vers l'infini. En outre, on constate une limite à l'effet d'amplification : au delà de deux fois le seuil, l'élasticité passe au-dessous de 2 et au-dessous de 1,5 après 3 fois le seuil, pour tendre vers 1. Les entreprises qui ont une production de masse auront une trésorerie beaucoup moins sensible que celles produisant à faible échelle, et ce, quelle que soit leur taille. La sensibilité de la trésorerie s'apprécie en fonction du niveau de production que l'on peut atteindre. Le commentaire des valeurs du coefficient d'élasticité se situant en deçà du seuil n'offre pas de difficulté particulière. On remarquera également que la valeur des seuils est proportionnelle au montant des charges de structure, **F**, respectives.

L'effet d'amplification du levier de trésorerie - d'exploitation et d'encaisse - par rapport au volume de l'activité est donc un phénomène relatif complexe, loin de justifier en toute circonstance un remplacement du travail par le capital qui, en gonflant les coûts fixes, fragilise la liquidité de l'entreprise sans apporter une amélioration systématique de la rentabilité.

[6] La « trésorerie nette après maintien de la valeur du capital » est également désignée par l'expression « trésorerie disponible » : voir Ternisien M., « L'importance du concept de trésorerie disponible », Revue Française de Comptabilité, mars 1995, pp. 72-77.

[7] Juhel J.-C., « Le seuil de solvabilité, instrument d'analyse financière et modèle de prise de décisions » Revue Française de Comptabilité, n° 262, décembre 1994, pp. 60-69.

En revanche, pour un niveau de charges de structure donné, et pour un niveau de production donné, la sensibilité de la trésorerie est directement conditionnée par la marge. Cette idée donne un sens complémentaire à la relation de l'élasticité précédemment établie. En systématisant cette observation on peut définir le concept de *« marge critique »*.

**b) L'élasticité de la trésorerie par rapport à la marge unitaire sur coûts variables : $E_{T/m}$ .**

Pour une production donnée, la marge unitaire sur coûts variables critique est celle pour laquelle il n'y a ni trésorerie virtuelle positive ni trésorerie virtuelle négative. Si la *« marge critique »* est notée m* , elle est égale à F/ Q.

En effet,

$$E_{T/m} = \frac{mQ}{mQ - F} = \frac{\mathbf{m}}{\mathbf{m - F/Q}}$$

et, E tend vers l'infini si ( mQ - F ) = 0 ; donc **m* = F / Q**

Le graphique suivant présente pour un niveau de production donnée, F/Q, c'est-à-dire la *marge critique* que par analogie à la terminologie précédente nous qualifierons *de liquidité à terme*, ou encore, si l'on ne retient pas dans « F » les charges calculées, la *marge critique de liquidité immédiate*. En deçà du seuil, m* , la trésorerie virtuelle est négative, au-delà elle est positive.

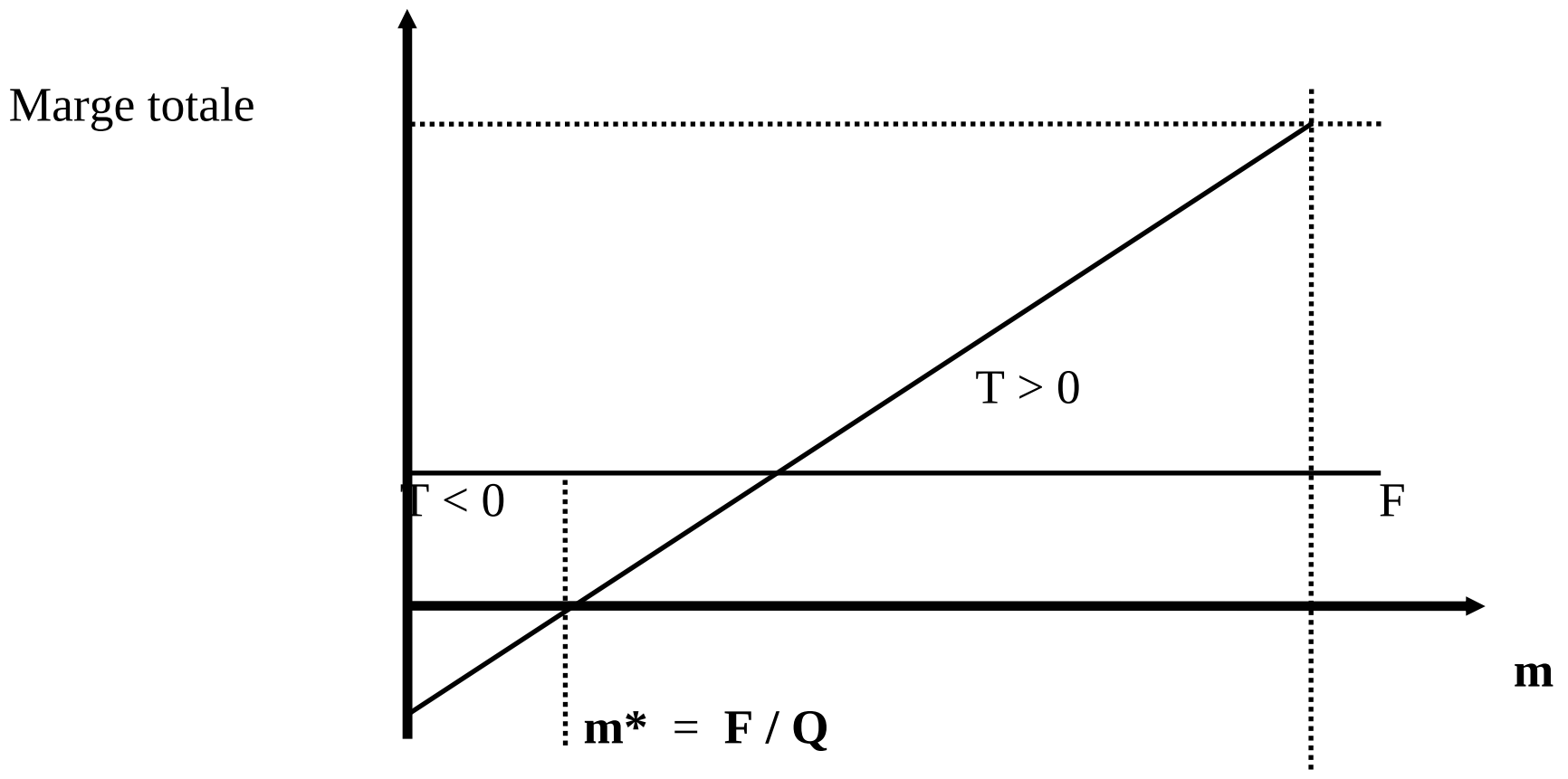

Le coefficient du levier de la liquidité à terme et celui du levier de la liquidité immédiate calculés par rapport à la marge ont un comportement identique à ceux calculés par rapport à la production, comme le montre le graphique suivant. Quelle que soit l'entreprise, l'effet d'amplification a la même configuration bien que lié à un couple spécifique - montant des coûts fixes et marge unitaire - caractéristique de la combinaison productive considérée. Au seuil de liquidité immédiate ou à terme, lorsque la marge totale couvre les coûts fixes afférents, l'élasticité tend vers l'infini.

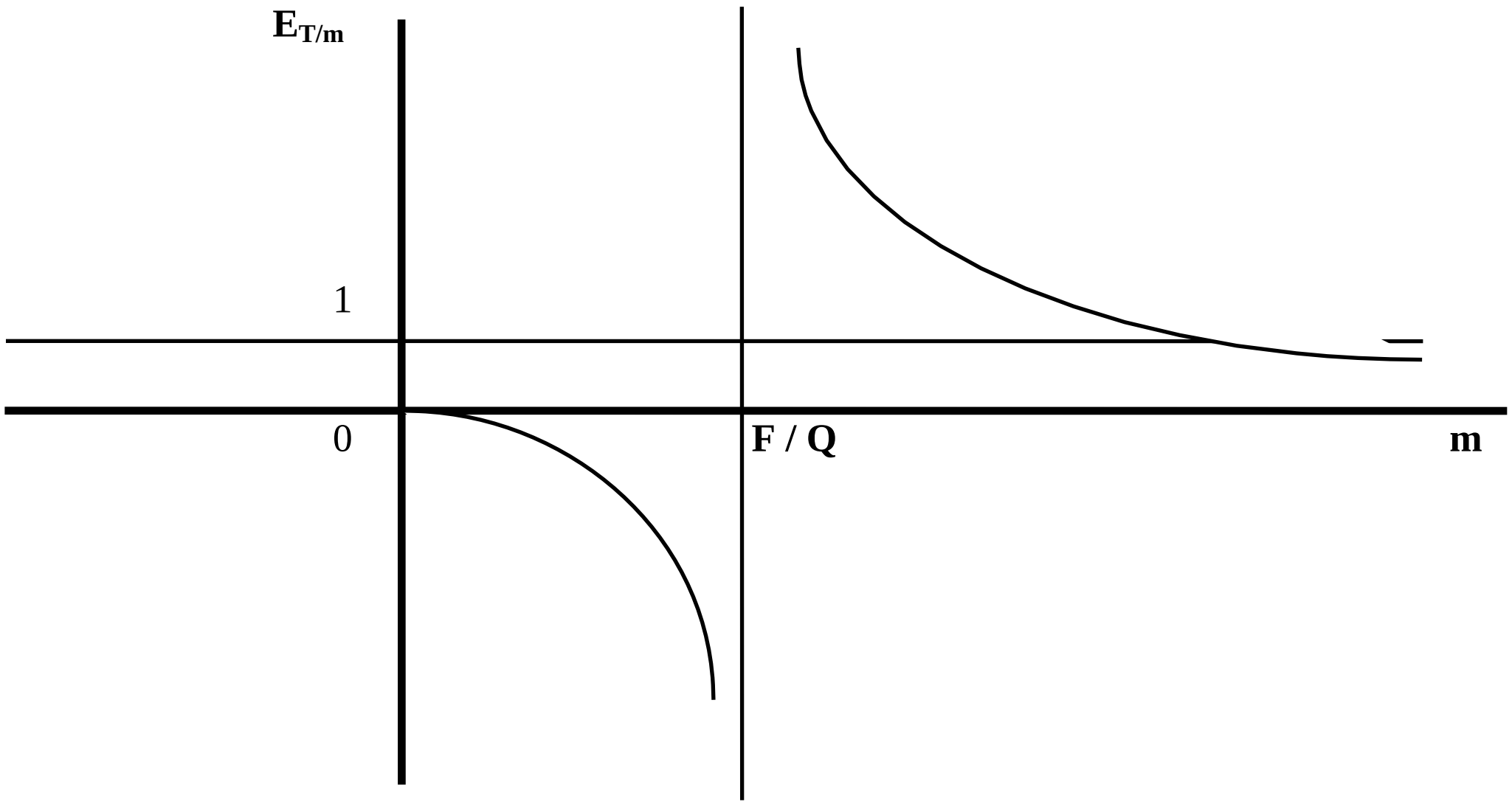

En effet, on peut écrire que lorsque la marge, m, augmente le rapport F/m tend vers zéro, et, $E_{T/m}$ tend vers 1 :

$$E_{T/m} = \frac{m\,Q}{m\,[Q - (F/m)]} \sim \frac{mQ}{mQ} = 1$$

Les valeurs que peut prendre l'élasticité en fonction de la marge sont résumée de la même façon que précédemment dans le tableau suivant :

| m | 0 | ½ F/Q | 2/3 F/Q | F/Q | F/Q | 2 F/Q | 3 F/Q | ∞ |
|---|---|---|---|---|---|---|---|---|
| **E** | 0 | -1 | -2 | - ∞ | + ∞ | 2 | 3/2 | 1 |

**L'effet de levier de trésorerie se compose, par conséquent, de quatre indicateurs de rupture de liquidité.** Le tableau suivant rassemble les indicateurs de rupture de la liquidité d'une structure d'exploitation :

| | **Production** | **Marge** |
|---|---|---|
| **Coûts fixes décaissables** | Seuil de liquidité immédiate | Marge critique de liquidité immédiate |
| **Coûts fixes totaux** | Seuil de liquidité à terme | Marge critique de liquidité à terme |

Ils ont une signification spécifique dans la mesure où, d'une part, ils prennent comme référence soit la production, soit la marge, et d'autre part, ils retiennent soit la totalité des coûts fixes soit les coûts fixes décaissables seulement. L'élasticité de la trésorerie immédiate mesure la sensibilité de « la solvabilité virtuelle » de la firme ; l'élasticité de la liquidité à terme mesure la sensibilité de « la trésorerie disponible » de fin de période. L'effet de levier de trésorerie est un indicateur stratégique pour les entreprises à faible capacité de production ou à marge réduite.

En combinant ces indicateurs on définit des **courbes d'indifférences de liquidité** déterminées par le **niveau des coûts fixes** pour lequel la trésorerie virtuelle est nulle. Elles associent un ensemble de combinaisons productives définies par leur marge ou leur capacité de production et **équivalente du point de vue de la liquidité.** Il est donc intéressant d'approfondir l'analyse des seuils de liquidité afin de tenter de définir **les caractéristiques des combinaisons productives à risque**.

Le graphique 1 suivant représente l’évolution des seuils de liquidité - immédiate ou à terme, selon le niveau de charges fixes retenues - en fonction de la conjonction du volume de production vendue et de la marge unitaire sur coûts variables, confrontée aux charges de structure caractéristiques de la combinaison productive concernée.

Ainsi, les coûts fixes totaux s’élevant à 8 000 000, on peut lire sur le graphique 1 que pour une marge de 8 le seuil de liquidité à terme est de 1 000 000 unités vendues, et pour une production de 2 400 000 unités la marge critique de liquidité à terme est de 3,33. Il existe donc des « *courbes d’indifférence de liquidité* » déterminées par le niveau des coûts fixes, pour lesquelles la trésorerie virtuelle est nulle, associant un panel de combinaisons productives définies par leur marge et leur capacité de production critiques et équivalentes du point de vue de la liquidité.

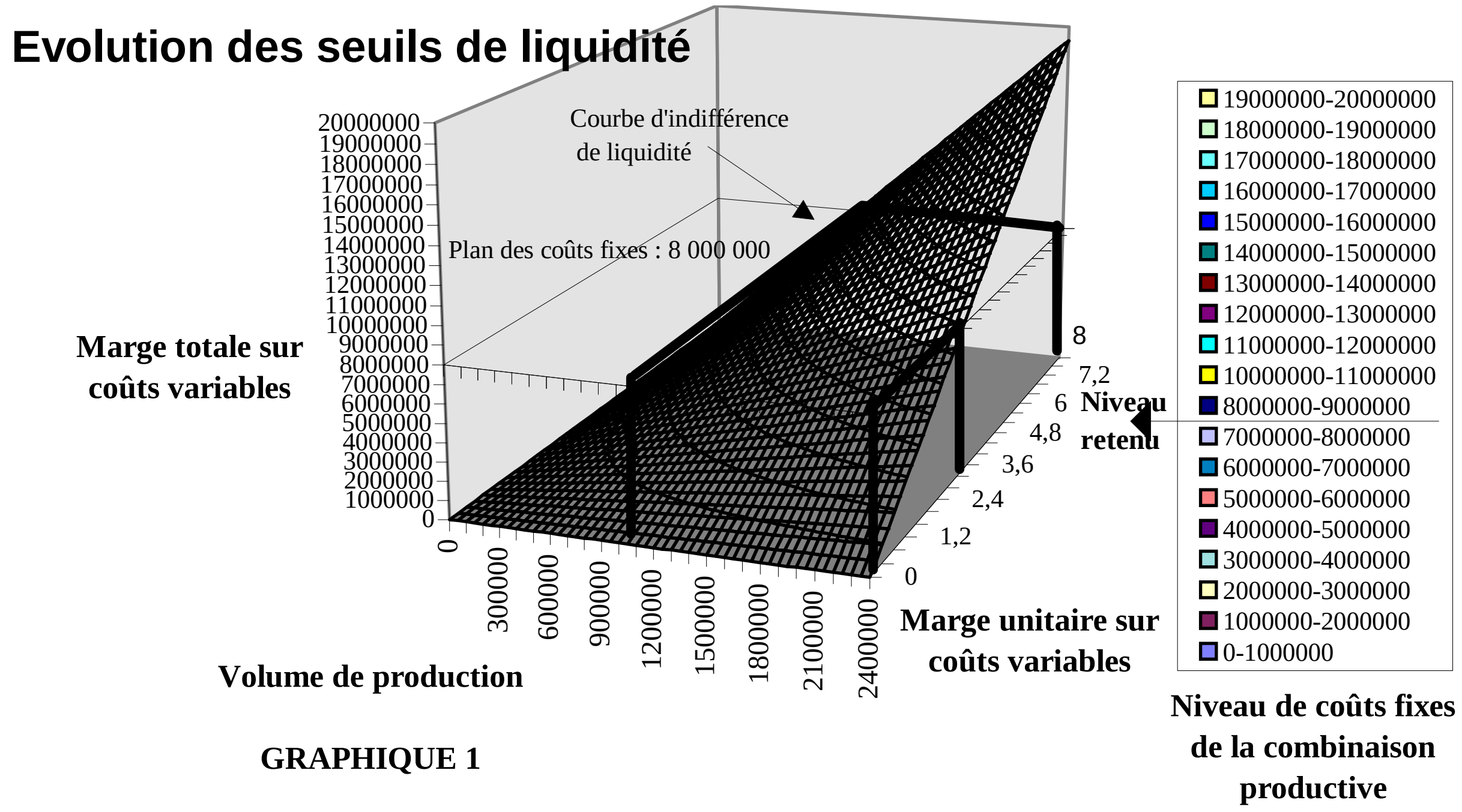

**GRAPHIQUE 1**

Le graphique 2 donne la projection sur un plan des courbes d’indifférences de liquidité correspondant aux paramètres retenus, et permet de considérer toutes les situations possibles.

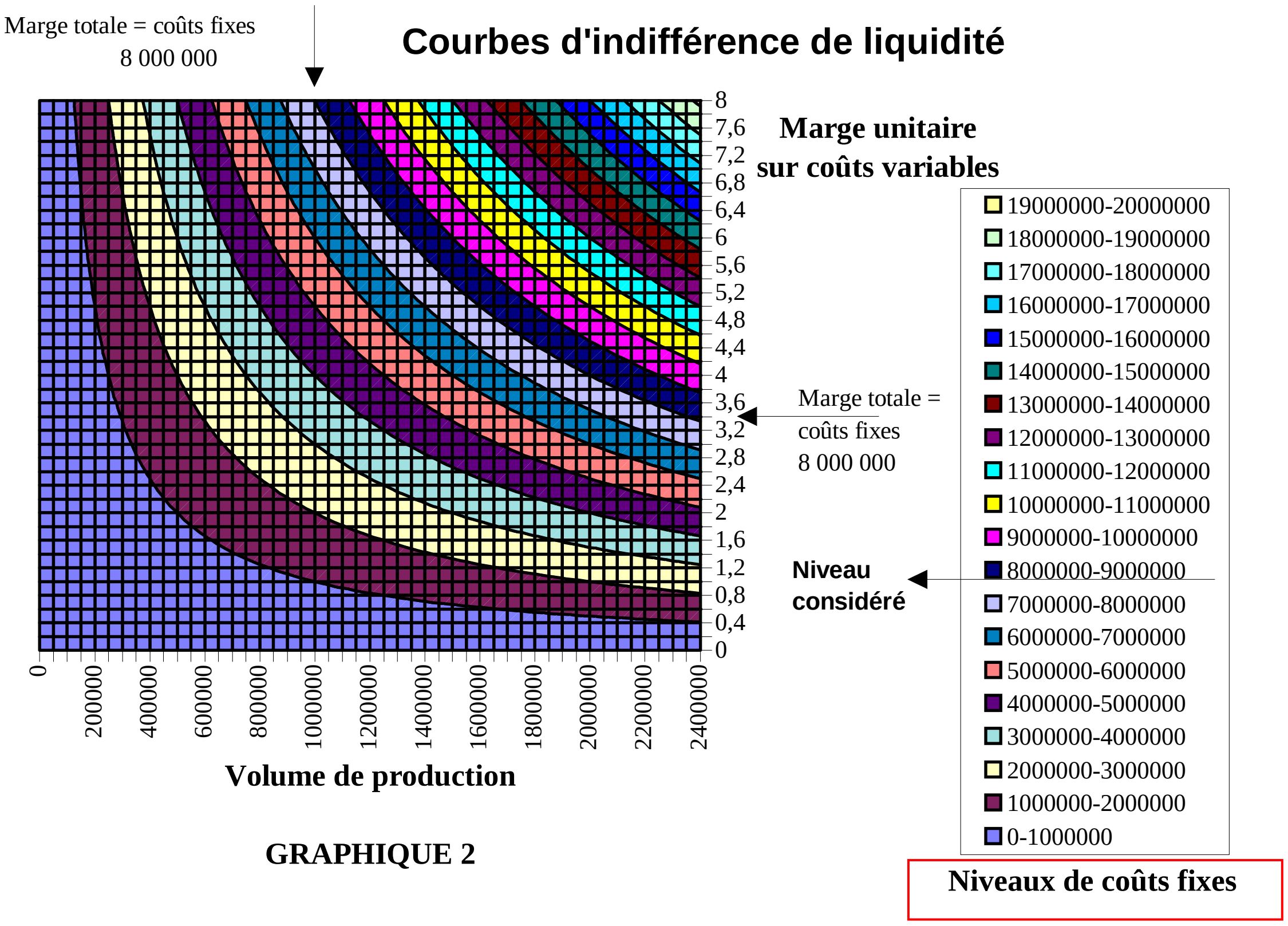

**GRAPHIQUE 2**

Ainsi, le risque d'insolvabilité dépend structurellement de la nature de la combinaison productive. L'effet de levier de trésorerie permet de mesurer la sensibilité de la trésorerie qui se trouve affectée par l'évolution des coûts fixes, de leur nature, de la marge unitaire sur coûts variables et du volume de production vendue.

L'effet de levier de trésorerie est un outil d'aide à la décision qui indique à l'entrepreneur les *conditions* dans lesquelles il peut *transformer son processus productif et développer son activité* en toute sécurité. Il permet de définir des courbes d'indifférence de liquidité qui orientent le choix de la combinaison productive vers des situations minimisant la sensibilité de la trésorerie aux aléas de l'activité.

Le choix de la structure du cycle d'exploitation est, comme le choix de la combinaison productive, stratégiquement déterminant dans la gestion de la trésorerie.

## II - Le choix de la structure du cycle d'exploitation.

La structure de l'exploitation est à l'origine de la formation du **flux de trésorerie disponible**[8]. Le flux de trésorerie disponible est source de valeur : il permet de limiter l'appel aux financements externes et de mieux rémunérer les actionnaires. La connaissance de son processus de formation est donc essentielle afin de l'utiliser comme outil stratégique. La théorie financière[9] analyse le flux de trésorerie disponible comme la confrontation entre la capacité d'autofinancement avant charges financières et l'investissement total. Le flux de trésorerie disponible résulte donc de l'efficacité de l'exploitation de l'entreprise et de sa politique d'investissement. Il obère à la fois sa croissance, son endettement, la rémunération du capital et ses réserves de liquidités. Autant dire la place cruciale de ce concept dans la gestion financière de l'entreprise et sa stratégie de développement[10]. L'entrepreneur poursuit **la maximisation du flux de trésorerie disponible** aussi bien en jouant sur **les facteurs de productivité** de son exploitation qu'en négociant ou en imposant à ses partenaires des **conditions d'échange** qui lui sont favorables[11].

[8] Free cash-flow ou « flux de trésorerie disponible ». Voir Jensen M.C. : « Agency Costs of Free Cash Flow, Corporate Finance and Takeovers », American Economic Review, vol. 76, 1986.

[9] Copeland T., Koller T. et Murrin J. : « La stratégie de la valeur », Interédition, Mac Kinsey 1991 - Rappaport A. : « Creating Shareolder Value : the New Standards of Business Performance », Free-Press, 1986.

Le flux de trésorerie disponible *après financement des investissements* se calcule généralement de la manière suivante :

Capacité d'autofinancement avant charges financières
- investissement net en besoin en fonds de roulement
**= Trésorerie d'exploitation**
- investissements nets (en capacité de production et actifs financiers)
**= Trésorerie disponible**

et recouvre donc :

la variation des liquidités et quasi-liquidités
+ la variation nette des capitaux propres
- versement des dividendes
+ la variation nette des dettes financières
- paiement des charges financières.

La couverture de l'investissement net en besoin en fonds de roulement par la capacité d'autofinancement ne peut véritablement s'apprécier qu'en termes de variation dans le temps, traduisant l'évolution de l'efficience de la combinaison productive et des rapports entre les parties prenantes à la production. A cet égard, comme on vient de le suggérer, la méthode des comptes de surplus constitue un instrument opérationnel d'analyse pertinent. La formation du « *surplus de trésorerie* » a pour origine la répartition des variations de flux d'exploitation d'un exercice à l'autre. De plus, cette approche permet d'intégrer spécifiquement l'investissement en besoins en fonds de roulement qui n'est autre que la résultante des décalages temporels entre flux d'exploitation et flux de trésorerie.

### A - La formation du surplus de trésorerie.

Le tableau de variation d'encaisse, le tableau de financement et autres tableaux de flux de trésorerie ne traduisent que très imparfaitement la structure des mouvements de fonds qui se produisent lors de la transformation des flux d'exploitation en flux de trésorerie. L'analyse en termes de *« surplus de trésorerie »* permet d'en éclairer les mécanismes fondamentaux d'apparition. L'évolution de la productivité, définie comme la combinaison de facteurs qui assure le maximum de résultat avec le minimum de moyen est la première source potentielle de fonds. Mais l'entreprise est une organisation socialisée. La méthode des comptes de surplus permet, en outre, de mettre en évidence le rôle des acteurs économiques dans la formation et l'évolution de la trésorerie virtuelle et dans les transferts potentiels de trésorerie.

---

[10] Hirigoyen G. : « Nouvelles approches du lien stratégie - finance », Revue Française de Gestion, janvier - février 1993, pp. 64 - 73. Ternisien M. : « L'importance du concept de flux de trésorerie disponible », Revue Française de Comptabilité, mars 1995, pp. 72-77.

[11] Juhel J.-C. : **Using surplus to measure the impact of cash transfers on free cash flow** - 20 pages - *Communication ''8th World Conference of the International Association for Accounting Education et and Research''* - Paris, 23-24-25 octobre 1997 - publié in « The changing world of accounting : global and regional issues » CDROM IAAER/AFC, Folio Transactive.

**a) L'origine du surplus de trésorerie virtuelle.**

Les performances de l'entreprise, c'est-à-dire son aptitude à créer de la valeur résulte de son efficacité à combiner les facteurs de production et à satisfaire la demande. La méthode des comptes de surplus est une analyse différentielle qui compare entre deux périodes la variation de la production et la variation des entrants utilisés pour produire (consommations intermédiaires et facteurs de production). La décomposition de chaque flux d'exploitation en « quantités - prix unitaire » permet de décrire l'origine du surplus de productivité. A cet égard, la valorisation des variations de flux exprimés en quantités, au prix de la période initiale élimine les distorsions apportées par une éventuelle variation de prix.

Lorsqu'une entreprise produit au cours d'une période davantage de biens ou de services, elle réalise un surplus de productivité si elle utilise des éléments de production (consommations intermédiaires et facteurs de production) augmentant dans une proportion moindre.

Soit une unité qui fournit i produits ou services (i = 1, 2, ... m) en utilisants j entrants de production (j = 1, 2, ...n). Les quantités produites notées « P » au prix unitaire noté « p » et les quantités utilisées notées « F » au prix unitaire noté « f » sont respectivement égales à :

- pour la période 1, $Pi$ et $Fj$ ;

- pour la période 2, $Pi + \Delta Pi$ et $Fj + \Delta Fj$ .

leur prix unitaire respectif s'écrit :

- pour la période 1, $pi$ et $fj$ ;

- pour la période 2, $pi + \Delta pi$ et $fj + \Delta fj$ .

Si l'on note R le résultat avant impôt de la période 1, le compte de production de la période 1 s'écrit :

$$\Sigma_i [ P_i . p_i ] = \Sigma_j [F_j . f_j] + R$$

Si l'on note $R + \Delta R$ le résultat avant impôt de la période 2, le compte de production de la période 2 s'écrit :

$$\Sigma_i [ (P_i + \Delta P_i) . (p_i + \Delta p_i) ] = \Sigma_j [ (F_j + \Delta F_j) . (f_j + \Delta f_j)] + [ R + \Delta R]$$

Par définition, le surplus dégagé par l'unité ou **surplus de productivité**, noté S est égal à :

$$S = \Sigma_i [ p_i . \Delta P_i ] - \Sigma_j [ f_j . \Delta F_j ]$$

Il est évident que ce surplus peut être négatif, ce qui signifie que l'entreprise connaît un affaiblissement de ces performances compensé par une évolution des prélèvements sur différents agents de la production.

A ce surplus de productivité ou surplus interne dégagé par l'unité, s'ajoute éventuellement un **surplus hérité** ou surplus externe qui résulte de la hausse des prix de vente, de la baisse des coûts de certains entrants ou encore de la diminution du résultat. Le surplus hérité noté S ' est égal à :

$$\mathbf{S' = \Sigma_j [ ( \Delta p_i ) . (P_i + \Delta P_i) ] + \Sigma_j [ (- \Delta f_j ) . (F_j + \Delta F_j) ] + (- \Delta R)}$$

Le **surplus global,** S + S ' est répartit entre les clients lorsqu'ils bénéficient d'une baisse des prix, les autres parties prenantes à la production lorsqu'il y a augmentation des prix d'entrants, et, enfin, à travers un accroissement du résultat, aux actionnaires, à l'entreprise elle-même - l'autofinancement - et à l'Etat ; il s'écrit :

$$\mathbf{S + S' = \Sigma_j [ ( - \Delta p_i ) . (P_i + \Delta P_i) ] + \Sigma_j [ ( \Delta f_j ) . (F_j + \Delta F_j) ] + \Delta R}$$

Appliquée à l'étude de la formation de la capacité d'autofinancement, cette analyse permet d'évaluer le surplus de trésorerie virtuelle de productivité. Distinguons dans le flux des charges exploitation noté $\Sigma_j [ F_j . f_j ]$ pour la première période et $\Sigma_j [ (F_j + \Delta F_j) . (f_j + \Delta f_j) ]$ pour la seconde, la partie du flux qui ne se transformera jamais en flux de trésorerie, car couvrant des charges non décaissables ( essentiellement les DAP ). Ce flux de charges non décaissables s'ajoute au flux « résultat » R et R + ΔR respectivement pour chaque période, pour former la capacité d'autofinancement de fin de période.

Si l'on note ce flux d'exploitation non décaissable $\Sigma_j [F_j^{dap} . f_j^{dap}]$ pour la période 1 et $\Sigma_j[(F_j^{dap} + \Delta F_j^{dap}) . (f_j^{dap} + \Delta f_j^{dap})]$ pour la période 2, la variation de la capacité d'autofinancement entre deux périodes successives, ou variation de la trésorerie virtuelle entre deux périodes résulte du jeu de l'évolution des trois facteurs suivant :

- d'abord, une variation potentielle de trésorerie engendrée par la productivité de l'entreprise que l'on peut appeler « variation de productivité de flux d'exploitation », ou **surplus de trésorerie virtuelle de productivité**, égal à :

$$\mathbf{\Sigma_i [ p_i . \Delta P_i ] - \Sigma_j [ f_j . \Delta F_j ]}$$

F ne désignant que des entrants source de décaissements. Ce surplus résulte de l'effort d'amélioration de la combinaison productive de l'entreprise. On compare le flux de liquidités engendré par l'accroissement de la production au flux de liquidités généré par l'augmentation des dépenses correspondantes. Si l'entreprise améliore son efficacité, elle accroît son flux de trésorerie potentiel, si son efficience s'affaiblit la variation du flux de trésorerie potentiel est négative. Elle prélèvera sur les différents agents de la production, lorsque le flux d'exploitation se transformera en flux de trésorerie, la trésorerie nécessaire. Ce sont évidemment les actionnaires qui couvriront le risque résiduel.

- ensuite, une variation positive potentielle de la trésorerie provenant des parties prenantes à la production que l'on nommera « variation héritée de flux d'exploitation », ou **surplus de trésorerie virtuelle hérité**, égal à :

$$\Sigma_j \left[ ( \Delta p_i ) . (P_i + \Delta P_i) \right] + \Sigma_j \left[ (- \Delta f_j ) . (F_j + \Delta F_j) \right]$$

Les mêmes remarques s'appliquent à F. Le transfert de trésorerie s'opérera, le moment venu, des agents de la production vers l'entreprise.

- enfin, une variation négative potentielle de la trésorerie provenant des autres parties prenantes à la production non concernée par le flux hérité, que l'on désignera par « variation transférée de flux d'exploitation », ou **surplus de trésorerie virtuelle transféré**, égal à :

$$\Sigma_j \left[ ( - \Delta p_i ) . (P_i + \Delta P_i) \right] + \Sigma_j \left[ ( \Delta f_j ) . (F_j + \Delta F_j) \right]$$

Les mêmes remarques s'appliquent naturellement à F. Là encore, le transfert de trésorerie s'opérera, le moment venu, de l'entreprise vers les agents de production.

Au total la valeur de la variation de la capacité d'autofinancement, ou « **surplus de trésorerie virtuelle** », qui en résulte s'élève à :

$$\Sigma_j (F_j^{dap} . \Delta f_j^{dap}) + (\Delta F_j^{dap} . \Delta f_j^{dap}) + (f_j^{dap} . \Delta F_j^{dap}) + \Delta R$$

**b) Les transferts potentiels de trésorerie.**

Le schéma suivant synthétise l'interférence des variations de flux d'exploitation précédemment décrites sur la variation de la trésorerie virtuelle de fin de période :

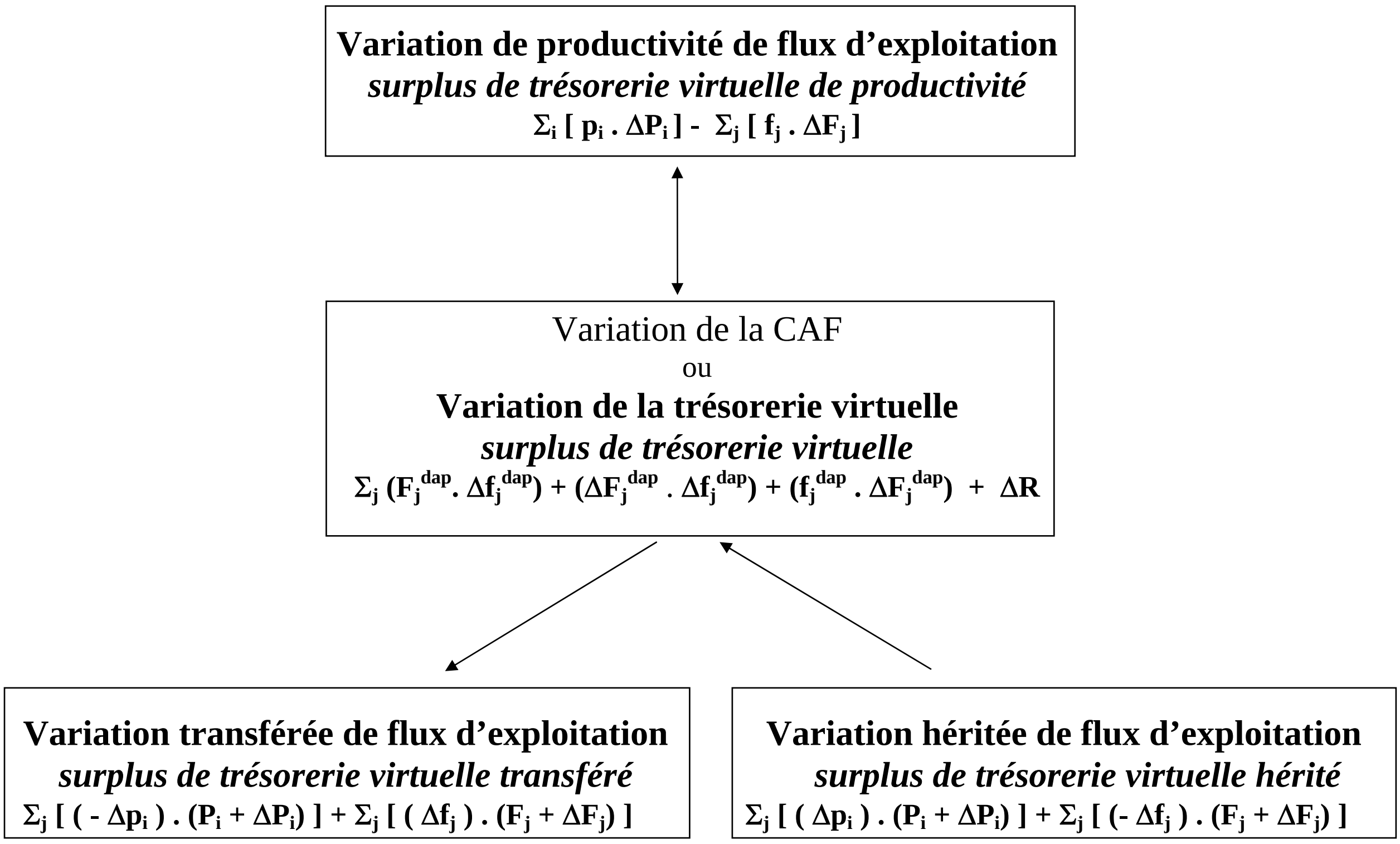

En intégrant l’incidence de l’imposition du résultat, on peut exprimer de la manière suivante le jeu des transferts potentiels de trésorerie entre les acteurs économiques et l’entreprise :

**1° Variation de productivité des flux d’exploitation** (amélioration ou affaiblissement des performances) provenant :

- d’encaissements potentiels de productivité (augmentation des quantités vendues à prix constant) ;
- de décaissements potentiels de productivité (augmentation des quantités d’entrants utilisés à coût constant).

**2° Variation transférée de flux d’exploitation vers** :

- les clients - encaissements potentiels réduits (diminution des prix sur les quantités vendues) ;
- les apporteurs d’entrants - décaissements potentiels accrus (augmentation des coûts sur les quantités d’entrants utilisés) ;
- à l’Etat - augmentation de l'impôt sur le résultat.

**3° Variation héritée de flux d’exploitation** :

- des clients - encaissements potentiels accrus (augmentation des prix sur les quantités vendues)
- des apporteurs d’entrants - décaissements potentiels réduits (diminution des coûts sur les quantités d’entrants utilisés)
- de l’Etat - baisse de l'impôt sur le résultat.

Ces variations de flux sont à l’origine des transferts de trésorerie et donc de risque d’insolvabilité entre agents de la production.

Avec une trésorerie nulle en début de période, la capacité d’autofinancement est la « trésorerie virtuelle » de fin de période. L’écart entre l’encaisse virtuelle et l’encaisse réelle tient aux décalages temporels entre les flux d’exploitation et les flux de trésorerie. La variation de la trésorerie d’exploitation traduit l’évolution de ces décalages puisqu’elle prend en compte l’investissement en besoin en fonds de roulement. Mais le besoin en fonds de roulement est un investissement particulier dans la mesure où il résulte de la différence entre emplois et ressources cycliques qui se renouvellent en permanence, de cycle d’exploitation en cycle d’exploitation. Un financement optimum ne peut que s’appuyer sur la connaissance de sa formation et de son évolution. Chaque décalage est le résultat de décisions de gestion. En effet, les transferts de fonds entre les parties prenantes à la production peuvent être retardés, parfois avancés selon les rapports entre partenaires. L’effet d’encaisse qui en résulte module la charge de financement de l’exploitation et par conséquent le risque de défaillance.

Pour connaître le mouvement de la trésorerie d’exploitation, il faut calculer les décalages entre flux potentiels et flux réels de trésorerie dus à la structure de l’exploitation. Pour chaque flux d’exploitation, on peut définir de manière classique des coefficients de retard ou d’avance dans leur écoulement à l’origine des flux de trésorerie. Ainsi, muni de ces coefficients de transfert on peut appréhender l’évolution de l’investissement en besoins de fonds de roulement au cours de la période.

**B - Analyse des décalages temporels : l’investissement net en besoin en fonds de roulement et la formation du surplus de trésorerie d’exploitation.**

En introduisant la notion de décalage temporel entre flux d'exploitation et flux de trésorerie on définit une nouvelle notion - *les transferts de trésorerie entre les parties prenantes à la production* - riche d'enseignement tant sur le plan de la gestion financière que sur celui du contrôle de gestion.

L'existence de délais de transfert provoque l'apparition d'un « flux d'encaisse net différé » sur l'exercice suivant. Les flux d'exploitation non encore encaissés ou décaissés à la fin de la période vont se transformer en flux de trésorerie au cours de la période suivante. Mais, au cours de cette deuxième période de nouveaux délais apparus diffèrent l'impact des flux d'exploitation sur la trésorerie. Ce double effet correspond à un désinvestissement et à un investissement en BFR. Le flux d'encaisse net différé constitue l'investissement net en besoin en fonds de roulement de l'exercice. Ajouté à la capacité d'autofinancement il formera la trésorerie d'exploitation. En revanche, pour décrire la formation du surplus de trésorerie d'exploitation de la période n, il faudra ajouter le flux d'encaisse net différé sur la période n-1 (ou investissement net en BFR de la période n-1), à la variation de la capacité d'autofinancement encaissée de la période n.

### a) Modalité de détermination des coefficients de transfert.

Représentatif des délais de transfert entre l'entreprise et les parties prenantes à la production, ces coefficients expriment les décalages entre flux d'exploitation et flux de trésorerie. La valeur relative du flux **k** stocké en fin de période se calcule en rapportant le montant du flux observé, $\mathbf{Fl_k}$, au stock, $\mathbf{S_k}$, qui lui est lié :

$$\text{Stock « } \mathbf{S_k} \text{ » / Flux « } \mathbf{Fl_k} \text{ »}$$

Le flux **k** peut être un flux de produit encaissable **i** ou un flux de charge décaissable **j** :

- $S_k$ représente $S_i$ ou $S_j$, c'est-à-dire un flux immobilisé ou stock de valeur en fin de période ;
- $Fl_k$ représente $Fl_i$ ou $Fl_j$, c'est-à-dire un flux observé au cours de la période afférent au stock $S_k$.

Donc, on peut écrire la valeur relative du flux immobilisé :

$$\mathbf{t_k = S_k / Fl_k}$$

Par conséquent, la valeur relative du flux k encaissé ou décaissé en fin de période 1 est égale à :

$$T_{k1} = 1 - t_{k1}$$

et, la valeur relative du flux k encaissé ou décaissé en fin de période 2, à :

$$T_{k2} = 1 - t_{k2}$$

Et la variation de la valeur relative du flux encaissé ou décaissé entre deux périodes successives se notera :

$$\mathbf{\Delta T_k = T_{k2} - T_{k1}}$$

et d'une manière générale, pour une période n :

$$\mathbf{\Delta T_{k(n)} = T_{k(n)} - T_{k(n-1)}}$$

La variation de la valeur relative du flux k peut porter sur un flux de produit encaissable i ou un flux de charge décaissable j, et s'écrire pour une période n :

$$\Delta T_{i(n)} = T_{i(n)} - T_{i(n-1)} \quad \text{ou} \quad \Delta T_{j(n)} = T_{j(n)} - T_{j(n-1)}$$

La variation de la capacité d'autofinancement **encaissée** est donc la combinaison d'un flux d'encaisse de productivité, d'un flux d'encaisse hérité et d'un flux d'encaisse transféré, c'est-à-dire de flux de trésorerie décalés par rapport aux flux d'exploitation correspondants.

Le **flux d'encaisse de productivité** est la différence entre l'encaissement de productivité et le décaissement de productivité.

D'abord, l'encaissement de productivité :

Si on note :

$\Delta P_i$, le supplément de production entre deux périodes successives, 1 et 2,

$p_i$, le prix unitaire des produits vendus au cours de la période 1,

$T_i$, le coefficient de décalage entre flux d'exploitation i et flux de trésorerie i au cours de la période 1,

on peut écrire que l'encaissement de productivité est :

$\Delta P_i \times p_i \times T_{i1}$ pour un produit

et $\Sigma_i [\Delta P_i \times p_i \times T_{i1}]$ pour i produits

Ensuite, le décaissement de productivité :

Si on note :

$\Delta F_j$, la variation de production entre deux périodes successives,

$f_j$, le coût unitaire des entrants achetés au cours de la période 1,

$T_{j1}$, le coefficient de décalage entre flux d'exploitation j et flux de trésorerie j au cours de la période 1,

on peut écrire que le décaissement de productivité est :

$\Delta F_j \times f_j \times T_{j1}$ pour un entrant,

et $\Sigma_j [\Delta F_j \times f_j \times T_{j1}]$ pour j entrants.

Au total **le flux d'encaisse de productivité**, noté $\Delta E_p$, gagnée à la fin de la période est égale à :

$$\mathbf{\Delta E_p = \Sigma_i [\Delta P_i \times p_i \times T_i] - \Sigma_j [\Delta F_j \times f_j \times T_j]}$$

Ce flux d'encaisse peut être négatif, ce qui signifie que l'entreprise connaît un affaiblissement de ces performances qui altèrent sa liquidité. Celle-ci devra être compensée par une évolution des prélèvements d'encaisse sur différents agents de la production.

Plus précisément, à côté du flux d'encaisse de productivité, l'entreprise peut soit transférer vers ses partenaires, soit hériter d'eux des flux d'encaisse :

$\alpha$) **Flux d'encaisse transféré** :

1 - encaissements réduits :

baisse de prix $\Sigma_i$ $[(P_i + \Delta P_i) \times \Delta p_i \times T_i]$ si $\Delta p_i <0$

allongement du transfert $\Sigma_i$ $[ (P_i + \Delta P_i) \times (p_i + \Delta p_i) \times \Delta T_i ]$ si $\Delta T_i < 0$

2 - décaissements accrus :

hausse des coûts $\Sigma_j$ $[ (F_j + \Delta F_j) \times \Delta f_j \times T_j ]$ si $\Delta f_j < 0$

raccourcissement du transfert $\Sigma_j$ $[ (F_j+ \Delta F_j) \times (f_j+ \Delta f_j) \times \Delta T_j ]$ si $\Delta T_j >0$

3 - augmentation de l'impôt sur le résultat :

si $\Delta R >0$

$\beta$) **Flux d'encaisse héritée** :

1 - encaissements accrus

hausse de prix $\Sigma_i$ $[(P_i + \Delta P_i) \times \Delta p_i \times T_i ]$ si $\Delta p_i > 0$

raccourcissement transfert $\Sigma_i$ $[(P_i + \Delta P_i) \times (p_i + \Delta p_i) \times \Delta T_i ]$ si $\Delta T_i >0$

2 - décaissements réduits

baisse des coûts $\Sigma_j$ $[ (F_j + \Delta F_j) \times \Delta f_j \times T_j ]$ si $\Delta f_j < 0$

allongement transfert $\Sigma_j$ $[ (F_j+ \Delta F_j) \times (f_j+ \Delta f_j) \times \Delta T_j ]$ si $\Delta T_j <0$

3 - baisse de l'impôt sur le résultat

si $\Delta R < 0$

Connaissant la valeur de la variation de la capacité d'autofinancement encaissée après impôt, ainsi calculée, analysons son contenu :

si l'on décompose la variation du flux des DAP = $\Sigma_j[(F_j^{dap} + \Delta F_j^{dap}) . (f_j^{dap} + \Delta f_j^{dap})]$ - $\Sigma_j [F_j^{dap} . f_j^{dap}]$, soit $\Sigma_j [(\Delta F_j \times f_j) + (\Delta f_j \times F_j) + ( \Delta F_j \times \Delta f_j)]$, où :

$\Delta F_j \times f_j$ exprime la valorisation de la variation des DAP en volume - *DAP (quantités)*

$\Delta f_j \times F_j$ exprime la valorisation de la variation des DAP en valeur - *DAP (prix)*

$\Delta F_j \times \Delta f_j$ exprime la valorisation de la variation des DAP en variations - *DAP (variations)*

et, si l'on note la variation du résultat après impôt encaissée, $\Delta R_E$*,
la variation de la capacité d'autofinancement après impôt et encaissée s'écrit :

$$\mathbf{= ( \Delta F_j \times f_j ) + (\Delta f_j \times F_j ) + (\Delta F_j \times \Delta f_j ) + \Delta R_E*}$$

On remarquera que les transferts de trésorerie n'interfèrent que sur la variation du résultat, $\Delta R_E$*, solde résiduel, et non pas sur les autres composantes de la trésorerie d'exploitation.

En outre, la différence entre la variation de la capacité d'autofinancement « comptable » et la variation de la capacité d'autofinancement « encaissée » d'une période à l'autre génère un flux différentiel, le « flux d'encaisse net différé » qui se reportera sur la période suivante. Ce flux d'encaisse net différé représente l'investissement net en besoin en fonds de roulement de la période.

### b) Le flux d'encaisse net différé.

Le flux d'encaisse net différé peut s'appréhender de plusieurs façons. Il se calcule indifféremment par différence entre la valeur comptable et la valeur encaissée de la **variation**, entre deux périodes successives, soit du résultat, soit de la capacité d'autofinancement, avant ou après impôt. Autrement dit, le facteur déterminant est l'évolution des décalages temporels entre flux d'exploitation et flux de trésorerie.

Le flux d'encaisse net différé affecte la variation de la trésorerie d'exploitation - ou surplus de trésorerie d'exploitation - de la période suivante. Le flux d'encaisse net différé sur la période n s'ajoute à la **variation** de la CAF encaissée de la période n+1, pour constituer le surplus de trésorerie d'exploitation de la période n+1. Si ce flux d'encaisse différé est positif, il y a désinvestissement, et le surplus de trésorerie d'exploitation augmentera d'autant. Si la variation est négative, elle s'imputera sur la variation de la capacité d'autofinancement, et réduira par conséquent la variation de la trésorerie d'exploitation.

Le tableau suivant présente l'analyse de la formation du surplus de trésorerie d'exploitation :

| Nature du flux d'encaisse | Mode calcul |
|---|---|
| **I - Flux d'encaisse de productivité** | 1 - 2 |
| 1 - Encaissements de productivité | $\Sigma_i [ \Delta P_i \times p_i \times T_i ]$ |
| 2 - Décaissements de productivité | $\Sigma_j [\Delta F_j \times f_j \times T_j ]$ |
| **II - Flux d'encaisse transférée** | 3 + 4 + 5 |
| 3 - Encaissements réduits | a + b |
| a) Baisse de prix ( $\Delta p_i$ <0 ) | $\Sigma_i [(P_i + \Delta P_i) \times \Delta p_i \times T_i ]$ |
| b) Allongement transfert ($\Delta T_i < 0$ ) | $\Sigma_i [(P_i + \Delta P_i) \times (p_i + \Delta p_i) \times \Delta T_i]$ |
| 4 - Décaissements accrus | c + d |
| c) Hausse de coûts ( $\Delta f_j < 0$ ) | $\Sigma_j [ (F_j + \Delta F_j) \times \Delta f_j \times T_j ]$ |
| d) Raccourcissement transfert ( $\Delta T_j >0$ ) | $\Sigma_j [(F_j+ \Delta F_j) \times (f_j+ \Delta f_j) \times \Delta T_j ]$ |
| 5 - Augmentation de l' I.S. ( si $\Delta R >0$ ) | |
| **III - Flux d'encaisse hérité** | 6 + 7 + 8 |
| 6 - Encaissements accrus | e + f |
| e) Hausse de prix ( $\Delta p_i > 0$ ) | $\Sigma_i [(P_i + \Delta P_i) \times \Delta p_i \times T_i ]$ |
| f) Raccourcissement transfert ( $\Delta T_i >0$ ) | $\Sigma_i [(P_i + \Delta P_i) \times (p_i + \Delta p_i) \times \Delta T_i$ |
| 7 - Décaissements réduits | g + h |
| g) Baisse de coûts ( $\Delta f_j < 0$ ) | $\Sigma_j [ (F_j + \Delta F_j) \times \Delta f_j \times T_j ]$ |
| h) Allongement transfert ( $\Delta T_j <0$ | $\Sigma_j [(F_j+ \Delta F_j) \times (f_j+ \Delta f_j) \times \Delta T_j ]$ |
| 8 - Baisse de l' I.S. ( si $\Delta R < 0$ ) | |
| **IV - Variation de la CAF encaissée** | 9 +10 + 11 + 12 = I + II + III |
| 9 - DAP (quantités) | $\Delta F_j \times f_j$ |
| 10 - DAP (prix) | $\Delta f_j \times F_j$ |
| 11 - DAP (variations) | $\Delta F_j \times \Delta f_j$ |
| 12 – $\Delta R$ après impôt encaissée | |
| | |
| **V - Flux d'encaisse net différé** | $\Delta R$ comptable - $\Delta R$ encaissée |
| | |
| **VI - Surplus de trésorerie d'exploitation** | IV ( n+1 ) + V ( n ) |

On observera sur le tableau que le flux d'encaisse net différé de la période n affecte par définition la variation de la capacité d'autofinancement encaissée de la période suivante, n+1. Le flux d'encaisse net différé résulte quant à lui de la différence entre les variations du résultat comptable et du résultat encaissée des périodes n-1 et n.

L'étude de l'origine de ces variations - variation de la capacité d'autofinancement encaissée et flux d'encaisse net différé - permettra de connaître la nature première de l'investissement en BFR, et, d'en apprécier les conditions de couverture par autofinancement avant tout autre projet d'investissement industriel ou financier. Le financement du cycle d'exploitation représente au cours de la vie de l'entreprise une des préoccupations essentielles de l'entrepreneur ayant le souci de maîtriser le risque de défaillance. La variation de la trésorerie d'exploitation joue, en effet, un rôle fondamental dans le processus de création de valeur.

A partir du tableau de financement on retrouve la valeur du surplus de trésorerie d'exploitation comme le montre le tableau suivant :

| Flux patrimoniaux | Période n | Période n+1 | Variation ( n+1 ) - ( n ) |
|---|---|---|---|
| **Capacité d'Autofinancement** | | | |
| **Variation du BFR** | | | |
| **Trésorerie d'exploitation** | | | **Surplus de trésorerie d'exploitation** |
| **Investissements nets** | | | |
| **Trésorerie disponible** | | | |
| **Variation de la dette financière** | | | |
| **Trésorerie disponible** **( après financement externe des investissements )** | | | |

En effet, la structure du tableau de financement est :

$$\Delta \mathbf{FR} - \Delta \mathbf{BFR} = \pm \Delta \mathbf{Trésorerie}.$$

Mais ici, si le flux d'encaisse différé est positif, il y a investissement net en BFR, et la trésorerie d'exploitation diminuera d'autant. Si le flux est négatif, le désinvestissement net augmentera la trésorerie d'exploitation. Le surplus de trésorerie d'exploitation sera la résultante, comme nous l'avons vu, de ces mouvements.

L'analyse proposée permet d'évaluer toute politique de restructuration du cycle d'exploitation en étudiant ses effets sur le flux de trésorerie disponible. Cette restructuration repose sur la modification des **facteurs de productivité** de l'exploitation et des **conditions d'échange** entre agents.

L'effet de levier de trésorerie et l'analyse du flux de trésorerie disponible en termes de surplus sont deux nouvelles approches essentielles du contrôle financier pour réaliser avec efficience les choix stratégiques que sont le choix de la combinaison productive et le choix de la structure du cycle d'exploitation.

## BIBLIOGRAPHIE :

Aoki M., Gustafsson B., et Williamson O.E., « The firm as a nexus of treaties », Sage Publications, London, 1990.

Boissonnade E., Palu J-C., « Délais de paiement : effets d'une réduction », Banque 1991, p. 908 et s.
Boulot J.-L., Cretal J.-P., Jolivet J. et Koskas S., « Analyse et contrôle des coûts », Publi-Union, 1986.
Chemillier Gendreau, « Financement des besoins d’exploitation des PME : enjeux et contraintes macro-économiques », Bull. mensuel du CIEG, novembre 1991.
Cohen J., Ewenczyk G., « L'analyste financier et le crédit commercial interentreprises », Analyse financière 1989.
Copeland T., Koller T. et Murrin J., « La stratégie de la valeur », Interédition, Mac Kinsey 1991.
Cooper R. et Kaplan R. S., « Profit priorities from activity-Based Costing », Harvard Business Review, mai 1991.
Darolles Y., Klopfer M., Pierre F. et Turq F., « La gestion financière », Publi-Union, 1986.
Depallens G. et Jobard J.-P., « Gestion financière de l’entreprise », Sirey, 10ème édition, 1990.
Fama E.F., « Agency problems and the theory of the firm », Journal of Political Economy, vol. 88, n° 2, avril 1980.
Fama E.F. et Jensen M.C., « Separation of ownership and control », Journal of Law and Economics, 26, 1983.
Fontanel B. « Stratégies financières et techniques comptables », Librairie Brunel, Vincennes, 1978.
Goffin R. « Principes de finance moderne », Economica, 1998.
Hirigoyen G., « Nouvelles approches du lien stratégie - finance », Revue Française de Gestion, janvier - février 1993, pp. 64 - 73.
Jensen M.C. ,« Agency Costs of Free Cash Flow, Corporate Finance and Takeovers », American Economic Review, vol. 76, 1986.
Juhel J.-C., « Le modèle du seuil de solvabilité, application pratique », 28 pages, revue « Echanges », mai 1995, n°112.
Lavoyer J.C., et Ternisien M., « Le tableau des flux de trésorerie », La Villeguérin Editions, Paris, 1989.
Levasseur M. et Quintart A., « Finance », 2ème édition, 1992.
Levy A., « Management financier », Economica, 1993.
Malo J.L., « Comptes de surplus », Encyclopédie de Gestion, Paris, Economica, 1989, p. 462.
Margerin J. et Ausset G., « Comptabilité analytique », Les Editions d’Organisation, 1993.
Massé P. et Bernard P., « Les dividendes du progrès », Le Seuil, Paris, 1969.
Monteil J. : « Les théories des surplus », Paris, Gauthier-Villars, 1966.
Pattison D. D. et Arendt C. G. : « Activity-Based Costing : It doesn’t work all the time », Management accounting , avril 1994.
Rappaport A., « Creating Shareolder Value : the New Standards of Business Performance », Free-Press, 1986.
Richard J., « Les tableaux de flux du cycle », Banque, n° 456 décembre 1985.
Roy J.L , « Un nouvel outil de la stratégie sociale : le surplus », Revue Française de Gestion, novembre 1977.
Ternisien M., « L’importance du concept de flux de trésorerie disponible », Revue Française de Comptabilité, mars 1995, pp. 72-77.
Ternisien M., « Comprendre l’entreprise par les flux » - La Villiguérin Editions, 1991.
Truel J.L., « Un outil de gestion à la française : les comptes de surplus », Harvard-L’Expansion, hiver 1979-80.
Van Horn J., « Gestion et politique financière », tome 2, Dunod, 1985.
Vassal J.C., « La méthode des surplus : applications à l’analyse du comportement des entreprises », Banque, 1972, n° 308 et 309.
Vernimmen P. « Finance d’entreprise », Dalloz,1998.
Wanieres A., « Le crédit interentreprises », Revue du financier, janvier 1991.
Weston J. F. et Brigham E. F., « La finance et le management de l’entreprise », CLM puli-union, 1973.
Williamson O.E. et Winter S.G., « The nature of the firm », Oxford University Press, 1991.